\documentclass[conference]{IEEEtran}
\makeatletter
\def\ps@headings{%
\def\@oddhead{\mbox{}\scriptsize\rightmark \hfil \thepage}%
\def\@evenhead{\scriptsize\thepage \hfil \leftmark\mbox{}}%
\def\@oddfoot{}%
\def\@evenfoot{}}
\makeatother
\usepackage{amsmath,amssymb,amsfonts}
\usepackage{algorithm}
\usepackage{algorithmic}
\usepackage{cite}
\usepackage{cprotect}
\usepackage{float}
\usepackage{graphicx}
\usepackage{listings}
\usepackage{tabularx}
\usepackage{textcomp}
\usepackage{xcolor}
\def\BibTeX{{\rm B\kern-.05em{\sc i\kern-.025em b}\kern-.08em
    T\kern-.1667em\lower.7ex\hbox{E}\kern-.125emX}}

\begin{document}
\title{False Sense of Security on Protected Wi-Fi Networks}

\author{
\IEEEauthorblockN{Yong Zhi Lim}
\IEEEauthorblockA{\textit{Information Systems Technology and Design}\\
Singapore University of Technology and Design}
\IEEEauthorblockA{yongzhi\_lim@mymail.sutd.edu.sg}
\and
\IEEEauthorblockN{Hazmei Bin Abdul Rahman\IEEEauthorrefmark{1}, Biplab Sikdar\IEEEauthorrefmark{2}}
\IEEEauthorblockA{\textit{Department of Electrical \& Computer Engineering}\\
National University of Singapore\\
\IEEEauthorrefmark{1}hazmei@u.nus.edu, \IEEEauthorrefmark{2}bsikdar@nus.edu.sg}
}

\maketitle

\begin{abstract}
The Wi-Fi technology (IEEE 802.11) was introduced in 1997. With the increasing use and deployment of such networks, their security has also attracted considerable attention. Current Wi-Fi networks use WPA2 (Wi-Fi Protected Access 2) for security (authentication and encryption) between access points and clients. According to the IEEE 802.11i-2004 standard, wireless networks secured with WPA2-PSK (Pre-Shared Key) are required to be protected with a passphrase between 8 to 63 ASCII characters. However, a poorly chosen passphrase significantly reduces the effectiveness of both WPA2 and WPA3-Personal Transition Mode. The objective of this paper is to empirically evaluate password choices in the wild and evaluate weakness in current common practices. We collected a total of 3,352 password hashes from Wi-Fi access points and determine the passphrases that were protecting them. We then analyze these passwords to investigate the impact of user's behavior and preference for convenience on passphrase strength in secured private Wi-Fi networks in Singapore. We characterized the predictability of passphrases that use the minimum required length of 8 numeric or alphanumeric characters, and/or symbols stipulated in wireless security standards, and the usage of default passwords, and found that 16 percent of the passwords show such behavior. Our results also indicate the prevalence of the use of default passwords by hardware manufacturers. We correlate our results with our findings and recommend methods that will improve the overall security and future of our Wi-Fi networks.
\end{abstract}


\section{Introduction}
Almost every aspect of our lives is associated with the usage of Internet. With the convenience of Wi-Fi, its usage as a medium of Internet access have increased drastically over the past two decades. The number of wireless broadband subscriptions in Singapore rose to 10.93 million in January 2018 with a wireless broadband population penetration rate of 193.9\% \cite{telecomstats}.

The WPA2-PSK security authentication mechanism is widely used in private Wi-Fi networks. It requires a passphrase with a minimum of 8 ASCII characters. Current Wi-Fi implementations on access points typically have no restriction on the selection of passphrases (e.g., selection of passphrases must contain at least an alphanumeric with special character). Many users tend to use easily identifiable passphrases (e.g., home/mobile phone numbers) or occasionally leave the default passphrase unchanged. In addition, some Wi-Fi manufacturers by default use 8 numeric characters for both the WPS (Wi-Fi Protected Setup) pin and WPA2 passphrase. 

The objective of this paper is to experimentally determine the prevalence of such practices in the wild. Our experiments also highlight various weak passphrase selection mechanisms that users often employ that decrease the effectiveness of both WPA2 and WPA3 encryption. A passphrase is defined as a password composed of a sequence of words \cite{correcthorsebatterystaple}. However, our results show that many private Wi-Fi networks can be compromised easily due to poor passphrase choices and when 8-digit passphrases are used, the passphrase may be broken within 10 minutes. In addition, our work highlights the use of default passphrases by multiple hardware vendors. 

The experimental results in this paper are based on two recent attack techniques. In May 2017, a novel attack technique known as \emph{Key Reinstallation Attack} (KRACK) was developed that achieves falsely secured information by manipulating and replaying cryptographic information in the 4-way handshake messages, which provides authentication for access into WPA2 encrypted networks \cite{krack}. In August 2018, an attack exploiting the \emph{Robust Security Network} (RSN) information element of a single \emph{Extensible Authentication Protocol over LAN} (EAPoL) frame was developed \cite{pmkid}. This attack allows an adversary to obtain the \emph{Pairwise Master Key Identity} (PMKID) without the need for a client. The \emph{PMKID} attack simplifies the process and allows performing an offline dictionary attack on the hash to gain the passphrase for the particular wireless network.

The WPA2 security standard has enjoyed its longevity in securing wireless networks for the last decade. As we move towards WPA3, this paper further addresses the need for stronger passphrases and identifies the extent to which the WPA2 attack (PMKID) affects currently deployed Wi-Fi routers and how targeted attacks against a badly implemented network (one that is secured with an 8 digit numeric passphrase) allows an adversary to easily gain access into the network. Furthermore, access to a WPA3-Personal network operating in transition mode is possible by performing a dictionary attack using the frames in the 4-way handshake \cite{dragonblood}.

We present the following contributions and findings, along with the practicality of performing such a study:
\begin{itemize}
    \item Collected PMKID hashes, discovered and identified which Wi-Fi manufacturers are affected by the WPA2 PMKID attack.
    \item Further understand how users cope with mechanisms in passphrase selection. %
    \item A novel approach to determine the strength of passphrases used for securing Wi-Fi networks to thwart common dictionary attacks and strengthen the overall landscape of password-based authentication.
\end{itemize}

The rest of the paper is organized as follows. Section 2 presents the related work and background material. The problem formulation is presented in Section 3 followed by the methodology in Section 4. The findings of the paper and a discussion on the results are presented in Section 5 and Section 6, respectively. Finally, the conclusions are presented in Section 7.

\section{Background and Related Work}
\subsection{Security Protocols for IEEE 802.11}

The Wired Equivalent Privacy (WEP) protocol was introduced in 1997 together with the first version of IEEE 802.11 standards to provide privacy in the wireless network. It uses the RC4 symmetric key encryption algorithm and it is shared among all stations connected to the Access Point (AP). It is well known that WEP has a critical security flaw as the WEP key is shared. Since the IV (Initialisation Vector) is 24 bits, it only provides $2^{24}$ combinations which leads to the possibility of having duplicated IVs in a short time. The IV is also transmitted in plain text which allows an adversary to easily inject and decode packets in the wireless network. 

Wi-Fi Protected Access (WPA) was developed in 2003 as a follow up to provide more sophisticated data encryption and better user authentication than WEP. It implements most of the IEEE 802.11i standard, especially the Temporal Key Integrity Protocol (TKIP). TKIP employs a per-packet key which dynamically generates a new 128-bit key for each packet thus preventing attacks that works on WEP encryption. However, a flaw in WPA encryption that relied on the older weaknesses in WEP and the limitations of the message integrity code hash function allows the retrieval of the keystream from short packets to be used for re-injection and packet spoofing.

Wi-Fi Protected Access 2 (WPA2) replaced WPA and is currently the widely used security encryption in IEEE 802.11 networks. It provides enhanced security in the MAC layer and includes mandatory support for CCM Mode Protocol (CCMP) which is an Advanced Encryption Standard (AES) based encryption mechanism. While WPA2 security encryption is more enhanced than WPA, it can be exploited by \emph{KRACK} and \emph{PMKID} attacks and its management frames are vulnerable to attacks as it is sent in plaintext. 
More recently, the WPA3 specification was released on 9 April 2018 \cite{wpa3-specification}. WPA3 is meant as a replacement to WPA2. It uses 128-bit encryption in personal mode and forward secrecy. WPA3 also replaces the Pre-Shared Key (PSK) exchange with Simultaneous Authentication of Equals (SAE) as defined in IEEE 802.11-2016 and introduces encryption for management frames. The Wi-Fi Alliance claims that WPA3 will mitigate security issues that are posed by weak passphrases. However, an adversary is still able to gain access into the network by bruteforcing his/her way into the network easily even if WPA3 encryption mitigate the 4-way handshake and PMKID attacks. This is done by performing an evil-twin and then requesting for exchange of the 4-way handshake via a downgrade to WPA2, making it invisible to the user. \cite{dragonblood}

\subsection{PMKID Attack}
In early 2018, a new attack on WPA2 security standard was accidentally discovered while Jens Steube, creator of hashcat, was looking for new ways to attack the WPA3 security standard \cite{pmkid}. While the WPA/WPA2 standards have already been compromised in the past using the man-in-the-middle approach, this attack does not require such approach thus being a client-less attack.

\begin{figure}[h]
\centering
\includegraphics[width=0.5\textwidth]{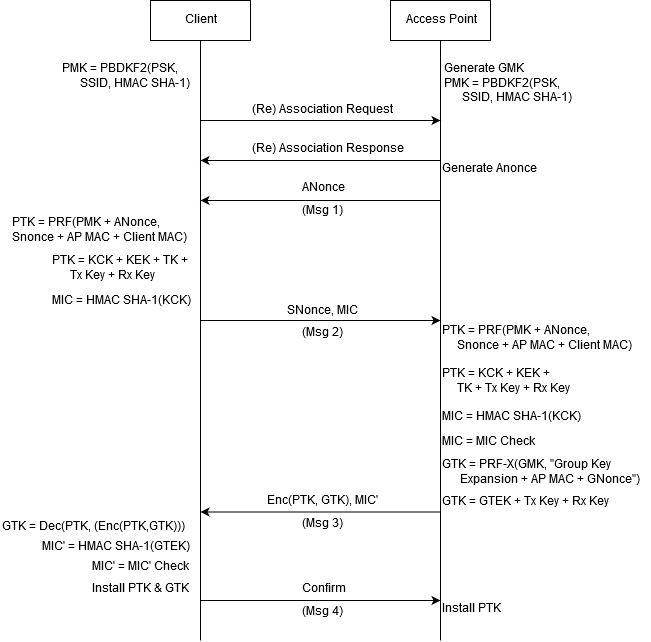}
\cprotect\caption{A diagram of the four-way handshake}
\label{fig:4-way_handshake}
\end{figure}

This attack exploits the Robust Security Network Information Element (RSN-IE) of a single Extensible Authentication Protocol over LAN (EAPoL) frame which is received during the authentication phase of the connection before the start of the four-way handshake, as shown in Fig.~\ref{fig:4-way_handshake} \cite{attackflow}. The RSN \emph{Pairwise Master Key Identification (PMKID)} can be retrieved under the WPA key data section as a hash value. This significantly accelerates the process of brute-forcing the network by communicating directly with the access point to retrieve the PMKID hash. The PMKID is the same as a regular EAPoL 4-way handshake and is computed using the HMAC-SHA1 function as follows: 
\begin{equation}
    PMK_{ID}=H(PMK,PMK_{Name}|MAC_{AP}|MAC_{STA})
\end{equation}
where PMK is the key to the function and the data part is a fixed string PMK Name, the MAC address of the AP, and the MAC address of the device trying to connect \cite{attackflow}.

For years, passwords relating to human recognition has always presented a key problem. %
As such, non-unified methods to truly determine the strength of passwords are varied and are instead compared with already known leaked databases such as \emph{haveibeenpwned.com}. Besides, access to a wireless AP can only be performed if only the passphrase is already known and not upon entry or generation by password managers.

\subsection{Password Strength}
\label{subsection:password-strength}
The US National Institute of Standards and Technology (NIST) defined the strength of a password by a function of length, complexity and unpredictability \cite{guessagain}. Due to human limited ability to memorize complex secrets, they often choose passwords that can be easily remembered or guessed. This leads to online services introducing rules in an effort to increase the complexity (e.g. composition rules).

There are 2 possible metrics to evaluate password strength: \textbf{Information entropy} and \textbf{Guessability}.
\begin{enumerate}
    \item \textbf{Information entropy}. Defined by Shannon as expected value in bits of information contained in a string. The entropy provides a lower bound on the expected number of guesses to find a text. The NIST publication (2006) uses entropy to represent the strength of a password.
    \item \textbf{Guessability}. This characterizes the time needed for an efficient password-cracking algorithm to discover a password. It is based on the human lack of creating random passwords with high entropy.
\end{enumerate}

However, many password based attacks associated with the usage are not affected by the complexity and length. Attacks such as keystroke logging and phishing attacks are out of the scope and will not be covered in this paper.

An effective measure of password strength is to calculate the entropy for it. Considering the worse case, the entropy of a password is at most $E = \log_2(\alpha^c)$, where $E$ = entropy (bits), $\alpha$ = number of possible ASCII characters and $c$ = number of characters in the password. This generates $\alpha^c$ possible passwords with $\log_2(\alpha^c)$, the number of bits in the entropy.

\begin{figure}[t!]
\centering
\includegraphics[width=0.5\textwidth]{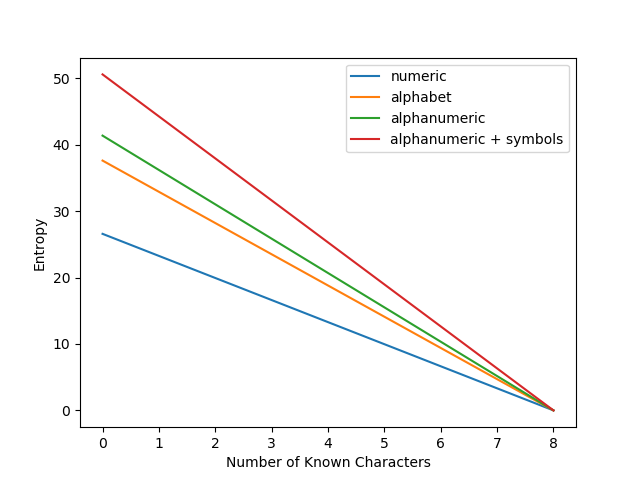}
\caption{Known characters vs entropy for an 8 character password}
\label{fig:char_entropy}
\end{figure}

However, should the best case be where one character of the password is known, the expression changes to $E_{best} = \log_2(\alpha^{c-1})$, which in turn creates the general equation of: 
\begin{equation}
E_{best} = \log_2(\alpha^{c-n}), \: \text{where $c \neq 0$}.
\end{equation}
where $n$ is not more than the number of characters in a password. An intelligent guess not only reduces the entropy but also effectively combat a perceived sense of security created by a user's weak password. This effectively reduces the entropy of the password and creates a false perception of the a strong password. Fig.~\ref{fig:char_entropy} shows the decreasing effects of entropy with each known character.

\subsection{Usage of Text-based Passwords}
Text-based password is a frequently used authentication method to access services or computer systems that are authorized for an individual. The most common and well known passwords such as ``\verb|password|'', ``\verb|123456|'', ``\verb|12345678|'', ``\verb|qwerty|'', ``\verb|abc123|'', ``\verb|111111|'',  ``\verb|monkey|'' are still widely used \cite{passwordisdead}.
Some statistics on the passwords used by users as at 2015 are presented below \cite{passwordisdead}. Password statistics include 91\% of passwords are one of the 1,000 most common passwords, 74\% of users use the same password for multiple accounts, 44\% of users change their password once, 21\% of users use passwords longer than 10 years, 54\% of users use 5 or fewer passwords during their entire life, the most common passwords are: ``password'' and ``123456'', more than 50\% of users forget their passwords and 7 in 10 users no longer trust passwords to protect their accounts.
Password theft statistics include 80\% of security incidents are caused by weak administrator passwords, 2 in 5 users had a password stolen in 2014, passwords with 6 lower-case letters are guessable in under 10 minutes and more than 378 million users become cyber crime victims annually.

In addition to the PMKID attack described earlier, other strategies may also be used to break Wi-Fi passphrases. A brute-force attack iterates through all possible character combinations until the correct password is found. This is done by comparing each password hash against the hash collected. A simple offline brute-force attack against an average MD5 password takes only around 3 days. With an additional 2 characters more it would take around 27 years \cite{passwordisdead}. This approach has a 100\% success rate and is only constrained by processing power and time.
A dictionary attack is a brute-force attack with a list of passwords (dictionary) instead of all character combinations. There is no guarantee to crack all passwords with this technique, but it significantly increases the speed and cracks most of the weaker passwords \cite{passwordisdead}. Context-specific dictionaries allows for very efficient password guessing.
Phishing attacks for Wi-Fi passphrases try to trick the user into giving the passphrase of the Wi-Fi network or the router. For example, this can be done by combining man-in-the-middle and denial-of-service attacks where the user is the directed to a webpage that looks similar to his/her router administrative page asking for the passphrase as the router requires a ``firmware update''.

Finally, the \emph{Key Reinstallation Attack} (KRACK) for Wi-Fi networks has been developed in \cite{krack}. A 4-way handshake is performed between clients and access point during Wi-Fi network authentication. The attack is based on a malicious actors sniffing the handshake to perform an offline bruteforce attack later on or view the traffic between the client and the access point. The malicious actor can either wait for a 4-way handshake to be performed or force a client to re-authenticate with the access point by sending disassociation/deauthentication frame. While bruteforcing the PMKID hash does not show a significant improvement compared to the 4-way handshake hash, the improvement it brings in retrieving the hash are significant. Clients are no longer required as the attacker directly communicates with the AP (client-less attack) and only the RSN-IE is required which is retrieved before the start of the 4-way handshake thus allowing for a greater range of attack.

\subsection{Related Work}
There have been investigations in the literature that have reported on the use of passwords in computer networks. An empirical study of password composition policies is presented in \cite{guessagain} while \cite{shay2016designing} evaluates the impact of fifteen composition policies on password strength and usability. The authors also evaluate  the resistance of passwords created under different conditions to guessing.
An evaluation of password strength meters used in many different applications and web environments is presented in \cite{high-impact_password}. A methodology for estimating password strengths is proposed in \cite{password_strength} and its effectiveness is evaluated in public datasets. A large-scale study of web passwords focusing on their re-use and strength is reported in \cite{large-scale_password_habits}.

The impact of human factors in password selection has been widely investigated in literature. The impact of human cognitive abilities in the selection of visual passwords has been investigated in \cite{influences_human_cognition}. The impact of user behavior and their cognitive load on the choice of passwords and their strength has been investigated in \cite{building_better_password, measuring_security_impacts_password}. Past research studies have also indicated that users write down their passwords or avoid the choice of higher-entropy passwords \cite{passwordsandpeople}. The ecological validity of password studies has been discussed in \cite{ecological_validity}.

In contrast to existing studies that focus on passwords in general computing environments, the focus of this paper is specifically on passwords used in Wi-Fi networks. Existing works do not consider wireless network passwords and in most cases, only a few hundred passwords are evaluated. In contrast, the results of the paper are based on more than a thousand passphrases collected from live and leaked networks \cite{wpa2-seclist}. Our results not only quantify the incidence of weak password choices in an ecologically valid study, they also demonstrate the impact of personality and cultural issues in cracking the passwords in Wi-Fi networks. Furthermore, our work also highlights the limitations of the human cognitive ability to create and remember difficult passwords.

\section{Problem Formulation}
In our study, our problem formulation restricts the Wireless Protected Access 2 - Pre-Shared Key (WPA2-PSK) to only 8 characters, considering the convenience of the user and the minimum required length to secure a Wi-Fi access point (AP). This constraint is practical as the human cognitive ability is limited in its ability to memorize complex secrets. 

The most common passphrase complexity consisting of lowercases, uppercases, numbers and symbols restricted to only 8 characters gives us the available number of possible combinations of $80^8 = 1.6777216e^{15}$.
However, restricting it only to numeric digits (0-9) will reduce the resulting possible combinations to $10^8 = 100,000,000$ which can be written for the general case as $10^n$, where $n$ represents the length of numeric digits required to be figured out in order to access the Wi-Fi AP. The entropy strength of a passphrase consisting of 8 numeric digits is approximately 26.58 bits, given $E = \log_2(10^{8})$. Considering the low entropy of these passphrases, these are trivial as they do not give a good sense of security. 

Note that we can lower the resulting complexity by performing an intelligent guess on the passphrase. The techniques used in this paper to reduce the entropy of a passphrase are listed below.

\subsection{Telephone Numbers}
According to Infocomm Media Development Authority of Singapore (IMDA), telephone numbers in Singapore are 8-digit long and typically starts with `3', `6', `8' or `9'. `3' represents a Voice over Internet Protocol (VoIP) number, `6' represents a residential number, while `8' and `9' represents a pre-paid and post-paid mobile number respectively \cite{imda-numbering}. Replacing $n$ from $10^n$ with `7' reduces the complexity to 10 million combinations and with the first number known as `3', `6', `8' or `9', the resulting complexity is simply $4*10^7 = 40,000,000$.

Furthermore, with the first digit known, we can further reduce the complexity if we deduce the area the user is in. The next 3 numbers in the phone number determines the area. For example, the number \verb|6750xxxx| shows that `6' is a residential line and `\verb|750|' is typically assigned to telephone numbers located in the north region of Singapore. This further reduces the required complexity to just $10^4 = 10,000$. This is a huge reduction from the required combinations in from $10^8$ and reduces the entropy strength to approximately 13.29 bits, given $E = \log_2(10^{4})$.

Naturally, for locations in the United States, telephone numbers can be specific to a state or city (e.g \verb|520| for Arizona), likewise in the United Kingdom (e.g \verb|0117| for Bistol or \verb|0161| for Manchester) which are easily georeferenced.

\subsection{Dates of Significance}
A user may have his or her event of significance (e.g., anniversaries or birth dates) set as a passphase. The most commonly used date format in Singapore is \verb|ddmmyyyy|, which results in exactly 8 numeric digits. For the current population, almost everyone is born between 1900 and 2019. With this known information, the resulting complexity is drastically reduced to 43,829 combinations or the number of days in 119 years. This is possible as such events of significance are written in the format \verb|ddmm19yy| and \verb|ddmm20yy|. Considering social engineering, this complexity can be further reduced if the day, month and/or year is known.

\subsection{Identification Numbers}
According the Section 5, National Registration Act, Cap. 201, Singapore issues the National Registration Identity Card (NRIC) and it is compulsory for Singaporean and foreign citizens who are permanent residents of Singapore \cite{nationalregistration}. However, the issue is the structure of the NRIC number which consists of 9 characters: a letter, 7 numbers then a checksum letter calculated using modulo 11. Similar to International Standard Book Numbers (ISBN), the algorithm to calculate the checksum letter was reverse engineered by Ngiam Shih Tung \cite{nric} and is calculated using the general formula:

\label{eq:nric}
\begin{align*}
    \mathtt{Checksum} = [\mathtt{Offset} + \: (d^1d^2d^3d^4d^5d^6d^7) \\ \cdot \; (2\,7\,6\,5\,4\,3\,2)] \; \mathtt{mod} \: 11 
\end{align*}
where $\mathtt{Offset}$ is given by 0 (with S or F prefix) or 4 (with T or G prefix).

\begin{center}
\begin{table}[h]
    \caption{Checksum Algorithm}
    \begin{tabularx}{\linewidth}{|X|m{0.7em}|m{0.7em}|m{0.7em}|m{0.7em}|m{0.7em}|m{0.7em}|m{0.7em}|m{0.7em}|m{0.7em}|m{0.7em}|m{0.7em}|}
        \hline
        \textbf{Checksum} & 10 & 9 & 8 & 7 & 6 & 5 & 4 & 3 & 2 & 1 & 0 \tabularnewline \hline
        \textbf{S, T Prefix} & A & B & C & D & E & F & G & H & I & Z & J \tabularnewline \hline
        \textbf{F, G Prefix} & K & L & M & N & P & Q & R & T & U & W & X \tabularnewline \hline
    \end{tabularx}
    \label{table:checksum_algorithm}
\end{table}
\end{center}

Again, by opting for the use of 7 digits and a letter (regardless of choice of the first or last checksum letter), the entire NRIC number sequence can be easily recovered.  The entropy strength of a passphrase consisting of a single letter (considering both upper and lowercases of the alphabet) with 7 numeric digits is approximately 28.95 bits, given $E = \log_2(54) + \log_2(10^{7})$. This can be compared to its use of Social Security Number (SSN) in the United States, where it contain 9 numbers instead.

\subsection{Default Passphrases}
We observed 7 Wi-Fi router manufacturers in our experiments (Asus, Aztech, D-Link, Linksys, Netgear, Prolink and TP-Link) and discovered that one of the manufacturers uses a default passphrase consisting of 8 digits the same passphrase for both WPA-PSK and WPS PIN. Table \ref{table:default_passphrases} shows some of the default PSK and WPS PIN that are used on different Wi-Fi models.\\

\begin{table}[h]
    \caption{Default Passphrases}
    \begin{tabularx}{\linewidth}{|l|l|X|X|}
        \hline
        \textbf{Manufacturer} & \textbf{Model} & \textbf{PSK} & \textbf{WPS PIN} \tabularnewline \hline \hline
        Aztech & DSL8800GR(S) & 0001543795 & 73930912 \tabularnewline \hline
        D-Link & DIR-878 & gcpjc44336 & 82200640 \tabularnewline \hline
        Linksys & EA8500 & vm0r76l6tq & 05855742 \tabularnewline \hline
        Linksys & WRT1900ACS & mnqe74ecv4 & 47062191 \tabularnewline \hline
        PROLiNK & PRC3801 & prolink12345 & \tabularnewline \hline
        TP-Link & Archer C9 & 25203738 & 25203738 \tabularnewline \hline
    \end{tabularx}
    \label{table:default_passphrases}
\end{table}

\section{Testing Methodology}
In the testing methodology, there are 2 phases involved; hash collection and hash bruteforcing. The following subsection provides the details of the 2 phases.

\subsection{Phase 1: Hash Collection}
In this phase, we utilize a \emph{Raspberry Pi 3 Model B+} as well as 2x \emph{D-Link DWA-137} Wi-Fi USB adapters (\emph{Ralink RT5372}) operating in monitor mode. With both \texttt{hcxdumptool} and \texttt{airodump-ng} running, we moved around several housing estates and the business district in Singapore.

\subsection{Phase 2: Hash Bruteforce}
For the second phase, we came up with several keyspaces to bruteforce through:
\begin{enumerate}
    \item Phone based digits starting with 3, 6, 8 and 9
    \item 8 digit characters (remaining that were not covered in the initial keyspace)
    \item NRIC - A letter followed by 7 digits
    \item NRIC - 7 digits with a last letter (checksum)
    \item NRIC
    \item Dictionary based on RockYou list
    \item Dictionary based on RockYou list with combination rule
\end{enumerate}

\texttt{hashcat} binaries was used for bruteforcing the PMKID hashes. During this phase, we discovered that 8 digit passphrase could be broken within 10 minutes. This could be further reduced to 60 seconds if we perform a targeted bruteforce with the first digit known (8, 6, or 9). 44.7\% of our results consists of phone based passphrases and if an adversary wants to perform a quick bruteforce, he/she can do so at an alarming rate of 1-5 seconds with a targeted bruteforce by limiting the keyspace to a specific region in Singapore (e.g. an adversary were to roam around central region of Singapore would try to use a keyspace of \verb|64500000| to \verb|64599999| as \verb|645| is the prefix for phone number in that area). %

\subsection{Formulation}
We assume the following parameters for this experimentation: 1) not taking keystroke dynamics (which takes into consideration of the user's input response time and accuracy), 2) the assumption of passphrases are primarily entered using a \emph{QWERTY} keyboard and 3) the numeric keypad is not taken into account for passphrase entry.

Based on our findings (shown later in Section \ref{section:findings}), most of the passphrases, be it generated by the user or device, consists of entry that can be mapped physically through the use of a physical or virtual keyboard. We then map this through a keyboard matrix as shown below:

\begin{figure}[H]
    \centering
    \small
    $
    \begin{bmatrix}
    \sim & ! & @ & \# & \$ & \% & \text{\textasciicircum} & \& & * & ( & ) & \_ & + & \\
     & Q & W & E & R & T & Y & U & I & O & P & \{ & \} & | \\
     & A & S & D & F & G & H & J & K & L & : & " &  & \\ 
     & Z & X & C & V & B & N & M & < & > & ? &  &  & 
    \end{bmatrix}
    $
    $
    \begin{bmatrix}
    ` & 1 & 2 & 3 & 4 & 5 & 6 & 7 & 8 & 9 & 0 & - & = & \\
     & q & w & e & r & t & y & u & i & o & p & [ & ] & $\textbackslash$ \\
     & a & s & d & f & g & h & j & k & l & ; & ' &  & \\ 
     & z & x & c & v & b & n & m & , & . & / &  &  & 
    \end{bmatrix}
    $
    \caption{A typical \emph{QWERTY} keyboard layout represented into a 3D matrix}
\end{figure}

Empty cells in the matrix exist provides an realistic gap between keys on the keyboard. As passphrase entry may usually consist of capitals and/or symbols, another layer is added on top of the usual matrix since the \emph{SHIFT} key is usually used as modifier for such an input.

By taking an user's input into account, we also can determine how a user transverse through the matrix sequentially by annotating a number in that particular cell in the matrix. For brevity, an example of a passphrase, \verb|password1| entered in the matrix is as shown below:

\begin{figure}[H]
    \centering
    
    \[
    \begin{bmatrix}
    0 & 9 & 0 & 0 & 0 & 0 \\ %
     & 0 & 5 & 0 & 7 & 0 \\ %
     & 2 & 3,4 & 8 & 0 & 0 \\ %
     & 0 & 0 & 0 & 0 & 0 \\ %
    \end{bmatrix}
    \begin{bmatrix}
    0 & 0 & 0 & 0 & 0 & 0 & 0 & \\
    0 & 0 & 0 & 6 & 1 & 0 & 0 & 0 \\
    0 & 0 & 0 & 0 & 0 & 0 &  &  \\ 
    0 & 0 & 0 & 0 & 0 &  &  & 
    \end{bmatrix}
    \]
    \caption{An example of a simple passphrase entered sequentially into the matrix separated into 2 halves of the keyboard}
\end{figure}

We can then calculate the distance between each key the user has input for the passphrase by using vectors. Since such an input can consist of two hands or thumbs representing a physical or virtual keyboard, the matrix can be then split into two halves. Any instance of capitals or symbols will result in another added dimension and is given in the general algorithm shown in Algorithm 1.

 \begin{algorithm}[h]
    \label{algorithm:vector-space}
    \caption{Determine Vector Space for Passphrase Input}
    \begin{algorithmic}[1]
        \renewcommand{\algorithmicrequire}{\textbf{Input:}}
        \renewcommand{\algorithmicensure}{\textbf{Output:}}
        \REQUIRE $(x_1 \cdot x_2 \cdots x_n), (y_1 \cdot y_2 \cdots y_n), (z_1 \cdot z_2 \cdots z_n)$
        \ENSURE $\vec{v}$
    
        \WHILE{passphrase entry is in progress}
            \FOR {$i = 1; i <$ \text{letters.length();} $i$++}
                \FOR {$j = 1; j <$ \text{letters.length();} $j$++}
                    \IF {lower case, number or symbol on $1^{st}$ layer}
                        \STATE obtain $(x_i, y_j, z_0)$
                        \STATE \textbf{break}
                    \ENDIF
                    \FOR {$k = 1; k <$ \text{letters.length();} $k$++}
                        \IF {upper case or symbol on $2^{nd}$ layer}
                            \STATE obtain $(x_i, y_j, z_k)$
                            \STATE \textbf{break}
                        \ENDIF
                    \STATE place sequence marker $(x_i, y_j, z_k)$ on map
                    \ENDFOR
                \ENDFOR
            \ENDFOR
            \STATE determine distance vector between \\ $(x_1 \cdot x_2 \cdots x_n), (y_1 \cdot y_2 \cdots y_n), (z_1 \cdot z_2 \cdots z_n)$
            \STATE sum both halves of vector maps
                \IF {$\vec{v} ==$ known passphrase}
                \STATE prompt user for likeness
                \ENDIF
        \ENDWHILE
        \RETURN $\vec{v}$
     \end{algorithmic}
 \end{algorithm}

The resultant will be a vector space and this accurately imprints how a passphrase is entered. For each character in the passphrase, a vector pointing towards the next character will always result in a positive vector. Naturally, as with inputs which are close to each other, distances between keys are short and this solves repeating sequences in passphrase entry. We argue that fundamentally the use of entropy (as explained in Section \ref{subsection:password-strength}) should not only be the given metric of a passphrase, but to also consider the sequence and the use of vectors as keyboard layouts, which has not changed since 1878. Besides, the consideration of having an additional dimension and actual sequence does not exist in \cite{chinesewebpasswords}, where its algorithm only identifies 3  patterns.

\section{Findings}
\label{section:findings}
We collected a total of 3,352 \emph{PMKID} hashes and define the findings as follows:
\begin{enumerate}
    \item Data Anonymity \\
    All data obtained were carried out in accordance with the Singapore's Cybersecurity laws as stipulated in Section 8A of the \textit{Computer Misuse and Cybersecurity Act (CMCA)}.
    \item Ethical Hacking \\
    The use of \emph{PMKID} mandates the execution of retrieving hashes from the AP. Only the hash retrieval and reversal process were carried out. No attempts of entry were made on any of the APs nor any deauthentication attacks were carried out.
    \item Exemption by Institutional Review Board \\
    Research and studies on normal human psychological responses and behaviours which are not designed or intended to study psychiatric or psychological disorders and which involve no more than minimal risk to the research subject exempts this research from the IRB.
\end{enumerate}

\subsection{Weaknesses in the Security Chain}
Upon inspecting a TP-Link Wi-Fi router (TL-WR710n), it was discovered that TP-Link uses the same 8-digit for both PSK and WPS pin by default. We further investigated newer TP-Link Wi-Fi routers (Archer C9 and TL-WR902AC) and our suspicion was correct. Users may likewise tend to use digits based on their cognitive ability, which is 8 digits long as their coping mechanism for passphrase selection.

Not to mention, TP-Link used the last 4 hex characters of the MAC address of their wireless hardware as their passphrase for connectivity to their routers \cite{tplink-mac} and limits the WPA2 passphrase to a maximum of 14 characters, as compared to the maximum allowable 63 in the IEEE 802.11i-2004 standard \cite{PassphrasesfordevicesecurityTPLinkSOHOCommunity}.

This should be classified as a security flaw as it gives a false sense of security for the user. A WPA2 or WPA3 (Transition Mode) network configured with 8-digit PSK can be easily bruteforced at a fast rate (maximum 10 minutes to cycle through the keyspace) compared to one that is alphanumeric. The use of strong security encryption is rendered useless in improving the security of the wireless network.

\section{Evaluations}
Out of the 3,352 hashes collected, we managed to find 536 passphrases ($\approx16\%$). We defined the different passphrase types as the following:

\begin{enumerate}
    \item (44.67\%) Phone-based passphrase
    \item (26.90\%) User-defined passphrase - simple combination of alphanumeric
    \item (15.23\%) Dictionary-based passphrase
    \item (7.11\%) Default (possible) passphrase
    \item (4.06\%) Date of Significance
    \item (2.03\%) NRIC
\end{enumerate}

On certain passphrases where an user uses phone number as the password, we noticed that there may be a correlation between the location and the passphrase. In other words, a Wi-Fi access point in a certain area may correspond its passphrase to its area code \verb|645XXXXX|. Using this information, an adversary would only need to bruteforce the remaining 5 digits in the passphrase thus, cutting down the duration of bruteforcing to a mere few seconds. %
Naturally, for use of passphrase in different geolocations, one can apply known passwords leaks in \cite{wpa2-seclist}, \cite{chinesewebpasswords} and \cite{zipfslaw}; where Chinese users are more likely to use only digits to construct their passwords, while English users prefer using letters. However, passphrases created for use in Wi-Fi networks mandate a minimum of 8 ASCII characters.

\begin{figure}[t!]
\centering
\includegraphics[scale=0.72]{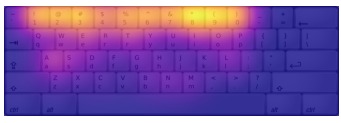}
\caption{Passphrase Heatmap}
\label{fig:complexity}
\end{figure}

\section{Conclusion and Future Work}

Although this research is focused solely within Singapore landscape, we noticed that users and some manufacturers uses simple passphrases (typically using only a string of digits or words) for the protection of the Wi-Fi network and there are no further safeguards in place to delay or prevent attacks as listed in this paper. Many routers are vulnerable to practical password recovery attack through the use of bruteforcing or dictionary attacks without the need of a connected client be it the network is operating in WPA2 or WPA3 Transition Mode. What is deemed secure to users to protect their Wi-Fi networks is now practically unsafe.

\newcommand{\BIBdecl}{\setlength{\itemsep}{0.25em}}
\bibliographystyle{IEEEtran}
\bibliography{bibliography}

\appendices
\setcounter{table}{0}
\renewcommand{\thetable}{A\arabic{table}}

\section{Recovered Passphases}
\label{app:passphrase}
Recovered passphrases and shown here are masked, and those that are duplicates or from dictionary attacks are omitted for security reasons.
\begin{table}[H]
    \caption{Date of Significance}
    \scriptsize
    \begin{tabularx}{\linewidth}{|X|X|X|X|}
        \hline
        0807xxxx & 1989xxxx & 0109xxxx & 0211xxxx \tabularnewline \hline
        0808xxxx & 1001xxxx & 2303xxxx & 2509xxxx \tabularnewline \hline
        1512xxxx & 1961xxxx & 1980xxxx & 2002xxxx \tabularnewline \hline
        2015xxxx & 2101xxxx & 2108xxxx & 2108xxxx \tabularnewline \hline
        2112xxxx & 2208xxxx & 2910xxxx & 3031xxxx \tabularnewline \hline
    \end{tabularx}
\end{table}

\begin{table}[H]
    \caption{Default}
    \scriptsize
    \begin{tabularx}{\linewidth}{|X|X|X|X|}
        \hline
        2380xxxx & 6759xxxx & 1311xxxx & 1582xxxx \tabularnewline \hline
        4344xxxx & 7490xxxx & 3346xxxx & 1327xxxx \tabularnewline \hline
        1833xxxx & 2491xxxx & 2502xxxx & 3205xxxx \tabularnewline \hline
        4612xxxx & 5518xxxx & & \tabularnewline \hline
    \end{tabularx}
\end{table}

\begin{table}[H]
    \caption{NRIC}
    \scriptsize
    \begin{tabularx}{\linewidth}{|X|X|X|X|}
        \hline
        S891xxxx & 8573xxxx & 8400xxxx & s890xxxx \tabularnewline \hline
        S852xxxx & s822xxxxx & s9144xxxx & S961xxxxx \tabularnewline \hline 
        t011xxxxx & & & \tabularnewline \hline
    \end{tabularx}
\end{table}

\begin{table}[H]
    \caption{Telephone Number}
    \scriptsize
    \begin{tabularx}{\linewidth}{|X|X|X|X|}
        \hline
        6457xxxx & 6552xxxx & 9299xxxx & 9641xxxx \tabularnewline \hline
        9774xxxx & 6250xxxx & 6286xxxx & 6364xxxx \tabularnewline \hline
        6366xxxx & 6453xxxx & 6457xxxx & 6457xxxx \tabularnewline \hline
        6457xxxx & 6457xxxx & 6463xxxx & 6554xxxx \tabularnewline \hline
        6565xxxx & 6899xxxx & 8127xxxx & 8163xxxx \tabularnewline \hline
        8181xxxx & 8181xxxx & 8282xxxx & 8322xxxx \tabularnewline \hline
        8323xxxx & 8383xxxx & 8488xxxx & 8589xxxx \tabularnewline \hline
        8765xxxx & 8781xxxx & 8888xxxx & 8939xxxx \tabularnewline \hline
        9007xxxx & 9008xxxx & 9026xxxx & 9027xxxx \tabularnewline \hline
        9049xxxx & 9090xxxx & 9109xxxx & 9125xxxx \tabularnewline \hline
        9129xxxx & 9170xxxx & 9180xxxx & 9183xxxx \tabularnewline \hline
        9186xxxx & 9226xxxx & 9236xxxx & 9271xxxx \tabularnewline \hline
        9272xxxx & 9339xxxx & 9350xxxx & 9368xxxx\tabularnewline \hline
        9452xxxx & 9456xxxx & 9466xxxx & 9632xxxx \tabularnewline \hline
        9653xxxx & 9665xxxx & 9667xxxx & 9669xxxx \tabularnewline \hline
        9674xxxx & 9677xxxx & 9692xxxx & 9732xxxx \tabularnewline \hline
        9744xxxx & 9745xxxx & 9749xxxx & 9750xxxx \tabularnewline \hline
        9782xxxx & 9785xxxx & 9794xxxx & 9824xxxx\tabularnewline \hline
        9848xxxx & 9858xxxx & 9859xxxx & 9880xxxx\tabularnewline \hline
        9891xxxx & 6452xxxx & 6453xxxx & 6458xxxx \tabularnewline \hline
        6636xxxx & 9889xxxx & 8461xxxx & 8765xxxx \tabularnewline \hline
        9733xxxx & 6909xxxx & 9232xxxx & 65674xxxxx \tabularnewline \hline
        6163xxxx & 6250xxxx       & 6254xxxx      & 6286xxxx \\ \hline
        6364xxxx & 6366xxxx       & 6452xxxx      & 6453xxxx \\ \hline
        6453xxxx & 6457xxxx       & 6457xxxx      & 6457xxxx \\ \hline
        6458xxxx & 6463xxxx       & 6552xxxx      & 6554xxxx \\ \hline
        6560xxxx & 6562xxxx       & 6563xxxx      & 6565xxxx \\ \hline
        6565xxxx & 6565xxxx       & 6565xxxx      & 6567xxxx \\ \hline
        6636xxxx & 6650xxxx       & 6698xxxx      & 6704xxxx \\ \hline
        6736xxxx & 6774xxxx       & 6776xxxx      & 6777xxxx \\ \hline
        6777xxxx & 6784xxxx       & 6814xxxx      & 6899xxxx \\ \hline
        6909xxxx & 6969xxxx       & 8102xxxx      & 8102xxxx \\ \hline
        8111xxxx & 8127xxxx       & 8139xxxx      & 8143xxxx \\ \hline
        8163xxxx & 8181xxxx       & 8188xxxx      & 8200xxxx \\ \hline
        8201xxxx & 8228xxxx       & 8228xxxx      & 8253xxxx \\ \hline
        8272xxxx & 8282xxxx       & 8304xxxx      & 8308xxxx \\ \hline
        8322xxxx & 8323xxxx       & 8383xxxx      & 8388xxxx \\ \hline
        8403xxxx & 8406xxxx       & 8458xxxx      & 8461xxxx \\ \hline
        8481xxxx & 8488xxxx       & 8589xxxx      & 8651xxxx \\ \hline
        8660xxxx & 8664xxxx       & 8666xxxx      & 8668xxxx \\ \hline
        8685xxxx & 8686xxxx       & 8699xxxx      & 8716xxxx \\ \hline
        8750xxxx & 8750xxxx       & 8765xxxx      & 8765xxxx \\ \hline
        8781xxxx & 8805xxxx       & 8813xxxx      & 8818xxxx \\ \hline
        8822xxxx & 8908xxxx       & 8939xxxx      & 9007xxxx \\ \hline
        9008xxxx & 9018xxxx       & 9026xxxx      & 9027xxxx \\ \hline
        9046xxxx & 9049xxxx       & 9068xxxx      & 9070xxxx \\ \hline
        9088xxxx & 9090xxxx       & 9109xxxx      & 9111xxxx \\ \hline
        9117xxxx & 9125xxxx       & 9129xxxx      & 9139xxxx \\ \hline
        9154xxxx & 9170xxxx       & 9179xxxx      & 9180xxxx \\ \hline
        9183xxxx & 9186xxxx       & 9226xxxx      & 9228xxxx \\ \hline
        9228xxxx & 9232xxxx       & 9234xxxx      & 9236xxxx \\ \hline
        9271xxxx & 9272xxxx       & 9299xxxx      & 9339xxxx \\ \hline
        9339xxxx & 9345xxxx       & 9350xxxx      & 9358xxxx \\ \hline
        9358xxxx & 9362xxxx       & 9366xxxx      & 9368xxxx \\ \hline
        9391xxxx & 9391xxxx       & 9398xxxx      & 9452xxxx \\ \hline
        9456xxxx & 9466xxxx       & 9479xxxx      & 9574xxxx \\ \hline
        9622xxxx & 9623xxxx       & 9632xxxx      & 9641xxxx \\ \hline
        9653xxxx & 9658xxxx       & 9665xxxx      & 9666xxxx \\ \hline
        9667xxxx & 9669xxxx       & 9669xxxx      & 9674xxxx \\ \hline
        9677xxxx & 9683xxxx       & 9683xxxx      & 9686xxxx \\ \hline
        9686xxxx & 9688xxxx       & 9692xxxx      & 9697xxxx \\ \hline
        9731xxxx & 9732xxxx       & 9733xxxx      & 9734xxxx \\ \hline
        9744xxxx & 9745xxxx       & 9745xxxx      & 9749xxxx \\ \hline
        9750xxxx & 9756xxxx       & 9762xxxx      & 9774xxxx \\ \hline
        9782xxxx & 9785xxxx       & 9794xxxx      & 9796xxxx \\ \hline
        9800xxxx & 9824xxxx       & 9825xxxx      & 9827xxxx \\ \hline
        9828xxxx & 9830xxxx       & 9830xxxx      & 9830xxxx \\ \hline
        9837xxxx & 9846xxxx       & 9847xxxx      & 9848xxxx \\ \hline
        9858xxxx & 9859xxxx       & 9866xxxx      & 9875xxxx \\ \hline
        9877xxxx & 9880xxxx       & 9889xxxx      & 9889xxxx \\ \hline
        9891xxxx &65674xxxx & &  \\ \hline
    \end{tabularx}
\end{table}

\begin{table}[H]
    \caption{User Defined}
    \scriptsize
    \begin{tabularx}{\linewidth}{|X|X|X|X|}
        \hline
        0000xxxx   & 0000xxxx   & 0012xxxx   & 0232xxxx     \\ \hline
        0544xxxx   & 0656xxxx   & 0804xxxx   & 0888xxxx     \\ \hline
        0909xxxx   & 0989xxxx   & 1003xxxx   & 1025xxxx     \\ \hline
        1111xxxx   & 1119xxxx   & 1122xxxx   & 1151xxxx     \\ \hline
        1234xxxx   & 1234xxxx   & 1234xxxx   & 1335xxxx     \\ \hline
        1413xxxx   & 1512xxxx   & 1683xxxx   & 1946xxxx     \\ \hline
        1961xxxx   & 2002xxxx   & 2108xxxx   & 3013xxxx     \\ \hline
        3031xxxx   & 3033xxxx   & 3059xxxx   & 3232xxxx     \\ \hline
        3240xxxx   & 3313xxxx   & 3318xxxx   & 3330xxxx     \\ \hline
        3333xxxx   & 3541xxxx   & 3685xxxx   & 4239xxxx     \\ \hline
        4280xxxx   & 4319xxxx   & 5019xxxx   & 5060xxxx     \\ \hline
        5240xxxx   & 5281xxxx   & 5295xxxx   & 5555xxxx     \\ \hline
        5637xxxx   & 5678xxxx   & 5707xxxx   & 5759xxxx     \\ \hline
        5807xxxx   & 5963xxxx   & 6666xxxx   & 7222xxxx     \\ \hline
        7275xxxx   & 7288xxxx   & 7318xxxx   & 7351xxxx     \\ \hline
        7954xxxx   & 8080xxxx   & 8888xxxx   & 12312xxxx    \\ \hline
        12345xxxx  & 32345xxxx  & 9876xxxxx  & 11112xxxxx   \\ \hline
        11223xxxxx & 12344xxxxx & 12345xxxxx & 12345xxxxx   \\ \hline
        12345xxxxx & 12345xxxxx & 12345xxxxx & 13579xxxxx   \\ \hline
        88798xxxxx & 12345xxxxx & 1234xxxx   & 12qwxxxx     \\ \hline
        1A2B3xxxxx & 1a2b3xxxxx & 7pvsxxxx   & a0909xxxx    \\ \hline
        a123xxxx   & a1234xxxx  & a1b2xxxx   & a1b2cxxxxx   \\ \hline
        abcdexxxxx & asdqxxxx   & Benjxxxx   & daniexxxx    \\ \hline
        lalaxxxx   & omnaxxxx   & Qwerxxxx   & qwertyxxxxxx \\ \hline
        ray1xxxx   & roygxxxx   & sfz8xxxx   & SJG3xxxx     \\ \hline
        sks1xxxx   &            &            &              \\ \hline
    \end{tabularx}
\end{table}

\lstdefinestyle{javascript}{
    basicstyle=\footnotesize\ttfamily,
    breaklines=true,
    frame=single,
    framesep=2mm
}

\section{Modifications to \texttt{validator.js}}
\begin{figure}[H]
    \centering
\begin{lstlisting}[style=javascript]
    psk: function(psk_obj, wl_unit){
        var psk_length = psk_obj.value.length;
        
        if(psk_length > 64){
            alert("<#JS_PSK64Hex#>");
            psk_obj.focus();
            psk_obj.select();
            return false;
        }
                        
        if(psk_length >= 8 && psk_length <= 63 && !this.string(psk_obj)){
            alert("<#JS_PSK64Hex#>");
            psk_obj.focus();
            psk_obj.select();
            return false;
        }
                        
        if(psk_length == 64 && !this.hex(psk_obj)){
            alert("<#JS_PSK64Hex#>");
            psk_obj.focus();
            psk_obj.select();                               
            return false;
        }
        
        var pass_exp = new RegExp(['^(?=.*[a-z])',
            '(?=.*[A-Z])',
            '(?=.*\\d)',
            '(?=.*[@$!%*?&])',
            '[A-Za-z\\d@$!%*?&]',
            '{8,64}$'].join(''));

        if(!psk_obj.value.match(pass_exp)){
            alert("Password does not meet complexity requirement!\nEnsure password length is between 8 to 63 characters and contains an upper, lower case, numeric and special character.");
            psk_obj.focus();
            psk_obj.select();
            return false;
        }

        return true;
    },
\end{lstlisting}
    \caption{Modifications to \texttt{validator.js}}
    \label{listing:validator}
\end{figure}


\end{document}